\documentclass{article}

\usepackage{PRIMEarxiv}

\usepackage[utf8]{inputenc} 
\usepackage[T1]{fontenc}    
\usepackage[hidelinks]{hyperref}       
\usepackage{url}            
\usepackage{booktabs}       
\usepackage{amsfonts}       
\usepackage{nicefrac}       
\usepackage{microtype}      
\usepackage{lipsum}
\usepackage{fancyhdr}       
\usepackage{graphicx}       
\graphicspath{{media/}}     

\usepackage[nohyperlinks, nolist]{acronym}
\usepackage{xspace}

\title{ANALYSE --- Learning to Attack Cyber-Physical Energy Systems With Intelligent Agents
}

\author{
  Thomas Wolgast, Astrid Nieße \\
  University of Oldenburg \\
  Oldenburg, Germany\\
  \texttt{firstname.lastname@uni-oldenburg.de} \\
   \And
  Nils Wenninghoff, Stephan Balduin, Eric Veith \\
  OFFIS - Institute for Information Technology \\
  Oldenburg, Germany \\
  \texttt{firstname.lastname@offis.de} \\
   \And
  Bastian Fraune, Torben Woltjen \\
  City University of Applied Sciences Bremen \\
  Bremen, Germany \\
  \texttt{firstname.lastname@hs-bremen.de} \\
}

\begin{document}
\maketitle

\newcommand{\eg}{e.g.\xspace}
\newcommand{\ie}{i.e.\xspace}
\newcommand{\etal}{et al.\xspace}

\newcommand{\smasher}{ANALYSE\xspace}
\newcommand{\palaestrai}{palaestrAI\xspace}
\newcommand{\mosaik}{mosaik\xspace}
\newcommand{\arsenai}{arsenAI\xspace}
\newcommand{\midas}{MIDAS\xspace}

\newcommand{\lastaccess}{2022-10-07\xspace}

\newcommand{\elasticfootnote}{\footnote{\url{https://www.elastic.co/elastic-stack/}, last access \lastaccess}\xspace}
\newcommand{\rettijfootnote}{\footnote{\url{https://gitlab.com/frihsb/rettij/}, last access \lastaccess}\xspace}
\newcommand{\elkstackfootnote}{\footnote{\url{ https://www.elastic.co/de/elastic-stack/}, last access \lastaccess}\xspace}
\newcommand{\mosaikfootnote}{\footnote{\url{www.offis.mosaik.de}, last access: \lastaccess}\xspace}
\newcommand{\midasfootnote}{\footnote{\url{https://gitlab.com/midas-mosaik/midas}, last access: \lastaccess}\xspace}
\newcommand{\pandapowerfootnote}{\footnote{\url{https://pandapower.readthedocs.io}, last access: \lastaccess}\xspace}

\begin{acronym}[JSONP]\itemsep0pt
	\acro{DER}{distributed energy resource}
	\acro{DSO}{distribution system operator}
	\acro{PV}{photovoltaic}
	\acro{RES}{renewable energy resources}
	\acro{TSO}{transmission system operator}
	\acro{WT}{wind turbine}
	\acro{RL}{reinforcement learning}
    \acro{MAC}{media access control}
	\acro{DRL}{deep reinforcement learning}
	\acro{AI}{artificial intelligence}
	\acro{ICT}{information and communications technology}
	\acro{OPF}{optimal power flow}
	\acro{CPS}{cyber-physical system}
	\acrodefplural{CPS}[CPS]{cyber-physical systems}
	\acro{CPES}{cyber-physical energy system}
	\acrodefplural{CPES}[CPES]{cyber-physical energy systems}
	\acro{DoE}{design-of-experiments}
	\acro{ARL}{adversarial resilience learning}
    \acro{SIEM}{security information and event management}
    \acro{YAML}{YAML Ain't Markup Language}
\end{acronym}

\begin{abstract}
The ongoing penetration of energy systems with \ac{ICT} and the introduction of new markets increase the potential for malicious or profit-driven attacks that endanger system stability.
To ensure security-of-supply, it is necessary to analyze such attacks and their underlying vulnerabilities, to develop countermeasures and improve system design. 
We propose ANALYSE, a machine-learning-based software suite to let learning agents autonomously find attacks in \aclp{CPES}, consisting of the power system, \ac{ICT}, and energy markets. 
ANALYSE is a modular, configurable, and self-documenting framework designed to find yet unknown attack types and to reproduce many known attack strategies in \aclp{CPES} from the scientific literature.

\end{abstract}


\begin{table*}[h]
\centering
\begin{tabular}{|l|p{6.5cm}|p{6.5cm}|}
\hline
\textbf{Nr.} & \textbf{Code metadata description} & \\ 
\hline
C1 & Current code version & 1.0 \\
\hline
C2 & Permanent link to code/repository used for this code version & \url{https://github.com/stbalduin/pyrate-analyse} \\
\hline
C4 & Legal Code License   & LGPL \\ 
\hline
C5 & Code versioning system used & git\\
\hline
C6 & Software code languages, tools, and services used & Python \\
\hline
C7 & Compilation requirements, operating environments \& dependencies & Can be run on most Unix systems like Linux and MacOS. Requires a Kubernetes Cluster with at least one node.\\
\hline
C8 & If available Link to developer documentation/manual & \url{https://github.com/stbalduin/pyrate-analyse/blob/main/README.md} \\
\hline
C9 & Support email for questions & palaestrai@offis.de \\
\hline
\end{tabular}
\caption{Code metadata}
\label{table:metadata} 
\end{table*}






\section{Motivation and Significance}

Energy systems worldwide are evolving into increasingly complex systems due to an increasing amount of \acf{ICT} added for monitoring and controlling, as well as the inclusion of energy markets. This results in faster time scales, smart controllable loads, and more automation and intelligence in the system. Despite the intended effects of controllability, adaptability, and cost efficiency of these digitalized systems, the resulting \acfp{CPES} exhibit new challenges for system operators with their increased complexity \cite{7980942}.
While system complexity increases, grid operators are under increasing economic pressure to maintain grid stability at minimal costs and investments. It must be ensured that no shortcuts are taken in terms of robustness.
\par 
Meanwhile, energy systems increasingly become targets for cyber attacks, terrorism, warfare, and other unwanted interventions from adversarial players. Recent examples are the ongoing attacks on the Ukrainian power system since 2017 and on the Nordstream gas pipelines in 2022.
Primarily due to the increasing complexity and interleaving of these interconnected systems, we expect an increasing potential for such attacks \cite{vei20}, with potentially catastrophic consequences. Besides attacks with malicious intentions, another possible motivation is to maximize profit by exploiting the flawed design of energy markets \cite{wol21}. However, the potential consequences can be the same, i.e., destabilizing interconnected energy systems.
\par 
\par

Therefore, an important research task is to develop methodologies and tools to identify systemic vulnerabilities, potential attack vectors, and unwanted destabilization incentives in existing and future \acp{CPES}. This way, preventive measures can be taken to remove vulnerabilities or to deploy countermeasures.
One major issue is that the still unknown attack vectors are the most relevant ones because no defense mechanism exists for them, similar to zero-day attacks in cyber security. While a growing body of literature investigates attacks on power systems, mainly rather predictable attacks or variants of known strategies are considered. 
However, given the systems-of-systems characteristics of \acp{CPES}, future attacks can exploit the system's increasing complexity and interleaving, making them impossible to foresee at design time with traditional approaches. 
\par 
One emerging approach to search for unknown attack possibilities is to place learning \textit{attacker} agents within the system, \eg \cite{wol21, ni18, yan17}. In literature, mostly \ac{DRL} agents are trained to maximize damage in the system, for example, by creating a blackout \cite{ni18}. In \ac{RL}, an agent uses trial-and-error to find actions in an environment that maximize a reward function \cite{sut18}. Consequently, learning attacker agents will autonomously develop strategies to attack the system if such strategies are possible and rewarded during training. From a research perspective, this allows us to extract new and unknown attack vectors and to stress-test a given \ac{CPES} in simulation. This is a prerequisite to developing defensive strategies in the future.
However, the learning agent can only utilize the degrees of freedom and information given. This is especially relevant in research, where often over-simplified scenarios are considered.
For example,  in current research, the different layers of a \ac{CPES}, \ie, physical component layer, \ac{ICT} layer, and function layer are typically modeled and evaluated separately. However, a too-limited action space may restrict the agent's ability to find new strategies, resulting in a false sense of security if no attack possibilities are found.
\par 
An \acl{AI} does not necessarily require a simplified system. Especially \ac{DRL} algorithms demonstrated how they can handle highly complex tasks, \eg, mastering the game of Go \cite{silver2016mastering}. 
Therefore, one next step in this research area is to move from simplified scenarios to more realistic and interconnected ones, yielding more realistic attack strategies. 
Applied to the energy system, \ac{ICT}, energy grids, and markets should be investigated more holistically as a \ac{CPES} to consider interdependencies.
\par 
In the following, we present \smasher (ANAlyzing compLex cYber-physical Systems wEaknesses), a tool-suite to analyze coupled power grids, \ac{ICT}, and market systems with learning agents for vulnerabilities and potential attack strategies. The main features of this software artifact are as follows:
\begin{itemize}
	\item \smasher is the first open-source tool-suite that combines power grid, energy market, and \ac{ICT} infrastructure into one coupled simulation to analyze them regarding vulnerabilities.
	
	\item It allows to place one or multiple learning (or non-learning) agents into the system that have access to sensors and actuators in all three domains. 
	\item \smasher is modular, configurable, self-documenting, and provides advanced data logging. 
\end{itemize}
Since \smasher utilizes a co-simulation framework, it is easily expandable, for example if alternative or additional domains need to be considered. 
This way, \smasher is designed to reproduce known attack strategies and scenarios from literature or to find novel more complex attack vectors over multiple domains.
\par 
In section~\ref{sec:description}, we provide an overview of \smasher and its components. Further, we present its key functionalities. In section~\ref{sec:example}, we provide a short illustrative example how the tool-suite works. In section~\ref{sec:impact}, we discuss the potential impact of \smasher before we conclude our work with section \ref{sec:conclusion}. 



\section{Software Description: \smasher}\label{sec:description}

\smasher describes the concept of multiple frameworks and modular simulators that work together as a tool-suite to analyze agent misbehavior in \ac{CPES}. 
We have developed multiple open-source tools that fulfill different sub-tasks towards that general goal but can also be used independently. This way, we achieve the best possible re-usability of the different software parts and a highly modular software architecture.
In the following, we will present the interplay of the \smasher components, the details of each component, and the overall functionalities. 

\subsection{Software Architecture}
\begin{figure}[p]
    \centering
    \includegraphics[width=0.9\linewidth]{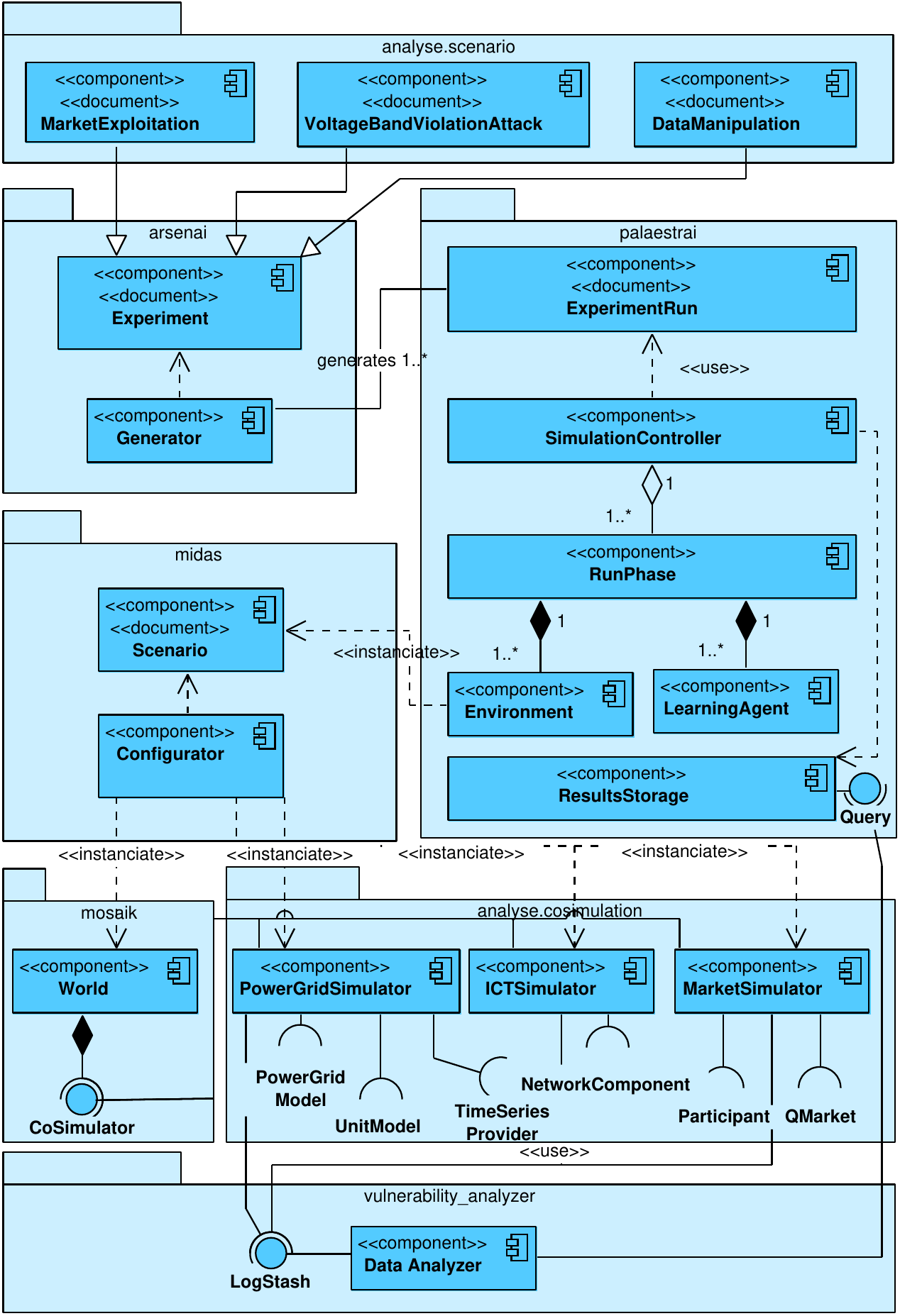}
    \caption{\smasher: interplay of system components.}
    \label{fig:system-components}
\end{figure}
The architecture and interplay of \smasher and its components are shown in Figure~\ref{fig:system-components}, which we describe in the following section from top to bottom.
The goal is to analyze attack scenarios in \ac{CPES}. The \emph{scenarios} are defined in a declarative way by YAML\footnote{\url{https://yaml.org/}, last access: 2022-12-22} files and are the starting points of \smasher experiments. They define which power grid to use, how many agents act in that environment, their objectives, etc.
The first step for an experiment is \arsenai, which translates the scenario file into multiple experiment runs by applying techniques from \ac{DoE}.
The experiment runs are performed by \palaestrai, an \ac{RL} framework that allows the training of one or multiple agents in an \ac{RL} environment.
To build modular \ac{RL} environments, we developed \midas, which assembles a single environment from multiple simulators -- in this case, power system simulator, market simulator, and \ac{ICT} simulator -- by using the co-simulation framework \mosaik.  
During the whole simulation and training, all relevant data is logged into a database. These data can later be analyzed regarding system vulnerabilities, noteworthy emerging agent strategies, potential countermeasures, etc.
The different components will be presented in more detail in the following sections. 


\subsubsection{Mosaik and MIDAS}

The open-source co-simulation framework mosaik\mosaikfootnote is, on a technical level, responsible for data exchange and synchronization of the different simulators \cite{steinbrink2019cpes}.
Mosaik follows a concept of so-called simulators and models.
A model can be anything, from a simple function to a complex simulation environment, while a simulator implements the mosaik API for that model.
A simulator can manage several models.
To allow different models to exchange data, they must be connected via mosaik.
The synchronization is done using individual but fixed time intervals for each simulator.

The more simulators and models are involved, the more complex the orchestration process of starting and connecting models and managing data flows becomes.
Therefore, we developed the open-source framework \midas\midasfootnote, which handles the assembly of mosaik scenarios and comes with some pre-configured smart grid simulators and scenarios. 
\midas also enables aggregating all mosaik simulators into an environment for \ac{RL} agents, defining actions, observations, and rewards.

\subsubsection{Power System Simulator}
\smasher uses the power grid simulator pandapower\pandapowerfootnote  to simulate a power grid with its topology and power flows \cite{thurner2018pandapower}.
To achieve more realistic behavior, we also added simulators for publicly available load profiles and a simulator for a \ac{PV} model that takes real weather data as input.
Since the \ac{PV} models depend on environmental conditions, we added a simulator for publicly
available weather data as well.
Although all those simulators are individual components of the mosaik scenario, we comprise them as the power system simulator in the following for simplicity.

\subsubsection{Reactive Power Market Simulator}
As a market system, we added a local reactive power market: In this type of markets, the grid operator procures reactive power from generators and other energy resources in the local system, mainly to perform voltage control \cite{wol22}. 
Reactive power markets are still under active research and were shown to be susceptible to profit-oriented attacks \cite{wol21, buc21} and the exercise of market power \cite{wol22, zho02}. We chose to consider a reactive power market because their local nature allows having clear system boundaries, in contrast to, \eg, the wholesale energy market \cite{wol21}.
\par 
Most reactive power market models from literature are based on solving an \ac{OPF}, \ie, the grid operator solves an optimization problem to determine the optimal reactive power procurement \cite{wol22}. That is problematic for \ac{DRL} since often thousands and millions of iterations are required.  
To reduce computational effort, we implemented a non-\ac{OPF}-based market, where the grid operator accepts the cheapest and most useful reactive power offers to perform voltage control in the system. 
\par 
To achieve competition, we also implemented some basic non-learning agents participating in the market. 
%
%
\subsubsection{rettij \ac{ICT} Simulator} 
The rettij\rettijfootnote \ac{ICT} simulator has been developed to support \ac{CPES} research regarding \ac{ICT} security \cite{Niehaus2022}. It provides a simple-to-use yet scalable network simulator that can be combined with co-simulation frameworks like mosaik. 
rettij makes use of the Linux network stack, as rettij's network components (i.e., switch, router, hosts) are based on containerization technology and use Kubernetes as an orchestrator. The network simulator's architecture allows to simulate realistic network traffic without the noise of management traffic that often occurs in plain container communication. rettij does not aim to simulate certain network technologies such as cellular or WiFi but chooses a more general approach where a normal \ac{MAC}-layer is simulated. However, the network channels can be throttled and delays and packet losses can be introduced to mimic such communication technologies. Network topologies are defined declaratively by YAML configuration files. It is possible to use any custom components within the simulation as long as it can run in a docker container. For \smasher, we implemented interfaces that allow for interaction between rettij and the agents to control and monitor parts of the network infrastructure. Sensors provide the ability to read network states, such as the utilization of network interfaces, while actuators offer the possibility to manipulate data in the network or restart targeted nodes. 
A more in-depth description of rettij and a standalone version of a co-simulation usage example can be found in \cite{Niehaus2022, AALE2020}.




\subsubsection{\palaestrai: Learning Agents Framework}

\palaestrai is a framework to train and evaluate learning agent systems. The difference to existing \ac{DRL} frameworks is its clear separation of agents, environment, and experimental framework. \palaestrai focuses on a reliable execution of experiments with learning agents in complex environments, including co-simulations. It allows the implementation of various learning agents (\ac{DRL}, neuroevolution, etc.) and a variety of environments. 

The \palaestrai subpackage \emph{arsenAI} parses experiment documents and uses statistical means to generate several experiment run definitions with specific parameter combinations defined in the experiment document.
Each experiment run is reproducible; feeding the same experiment run document to different \palaestrai instances will produce the same experimentation result each time, which helps countering the AI reproducibility crisis \cite{hutson2018artificial,gundersen2018reproducible,haibe2020transparency}.
The key design goal of \palaestrai was to facilitate complex experiments, where agents can act in different, co-simulated simulators that form an \ac{RL} environment while learning from the experiences gathered from all simulators.


\subsection{Software Functionalities}\label{sec:functionalities}

The core of \smasher is the co-simulation ability gained from the mosaik interface. It interfaces to separate simulators for power grid, market, and \ac{ICT}. Through \palaestrai's approach, these three simulators can be treated as one large \ac{RL} environment. For example, an agent can observe the power grid's current state and decide to bid on the market based on this observation. The bid is then communicated to the market by \ac{ICT} components. The goal is to simulate potential attack vectors for analysis, such as manipulating the power grid through malicious assets, shutting an actor out of the market with denial-of-service attacks (DoS), or gaming the market through an agent that controls multiple assets. More importantly, \smasher allows learning agents to discover weaknesses that emerge through the interconnection of three complex systems.

To investigate and analyze a variety of potential attack scenarios, we emphasize the following key features:

\paragraph{Co-Simulation and Modularity} 
\smasher uses \mosaik to aggregate multiple independently developed simulators to create a single \ac{RL} environment. This results in a modular tool where simulators can be added, removed, or exchanged easily. For example, a simulator of the gas system or the balancing power market could be added.

\paragraph{Self-Documentation with Run Files}
Inherited from \palaestrai, experiments in \smasher are defined with run files. Run files are YAML files that define which \ac{RL} algorithm, environment, actuators, sensors, objectives, etc. are used for the experiment. This improves documentation of large-scale experimentation because the exact experiment definition for every experiment is always automatically stored. 
Further, this allows us to investigate diverse variants of one environment. Usually, in \ac{RL} research, the environment is a fixed benchmark environment. However, in applied \ac{RL},  we often want to investigate multiple variants of the environment, \eg, to explore the consequences of different design decisions, like controller design, market design, etc.

\paragraph{Logging and Evaluation}
To evaluate experiments, we designed a custom vulnerability-analyzer component. Its core concept stems from cyber-security, where log data from different sources is collected, harmonized, and shipped to a \ac{SIEM}. The \ac{SIEM} allows analyzing the data and correlations of events. To easier collect logs, the \smasher components create event logs and simulation status logs. Those logs are shipped to an Elasticsearch database and can be visualized with Elastic's Kibana interface. This typical Elastic stack\elkstackfootnote architecture allows efficient thread hunting \cite{elk-threat-hunting} for \ac{ICT}-security and is here used to approach experiment evaluation analysis. This flexible tool chain enables integrating more data towards a big data platform to detect interdependent attacks.

\section{Illustrative Example}\label{sec:example}
\begin{figure}[h]
    \centering
    \includegraphics[width=0.65\linewidth]{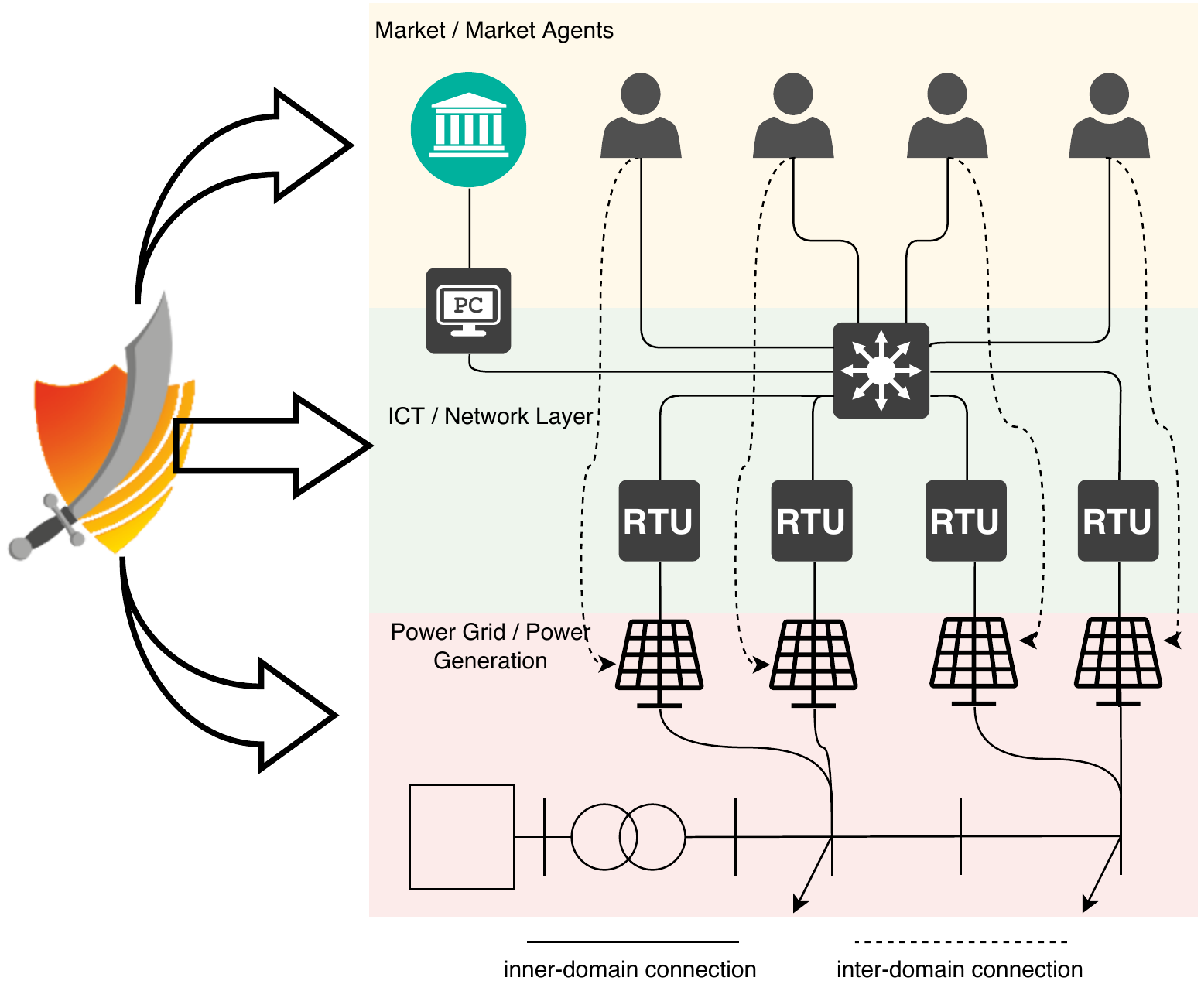}
    \caption{\smasher: Illustrative example consisting of a power grid, a local market, market participants, the communication network, and the attacker agent.}
    \label{fig:example}
\end{figure}

The following example demonstrates the capabilities of our software tool-suite \smasher. Figure~\ref{fig:example} shows a simple scenario with a small power grid, a local reactive power market, and the underlying \ac{ICT}. The power grid consists of a 4-bus network with two \ac{PV}-panels at bus~3 and 4, respectively, managed by four agents. Each agent can offer the flexibility of its \ac{PV} on the market. 

The market operator communicates with the market agents by using an \ac{ICT} network. To simplify the \ac{ICT} network, we have connected them to one switch. The \ac{ICT} network is also used by the market operator to communicate to the \acp{PV} how much reactive power they should deliver, \ie, how much power was procured on the market.

\begin{figure}[h]
    \includegraphics[width=0.9\linewidth]{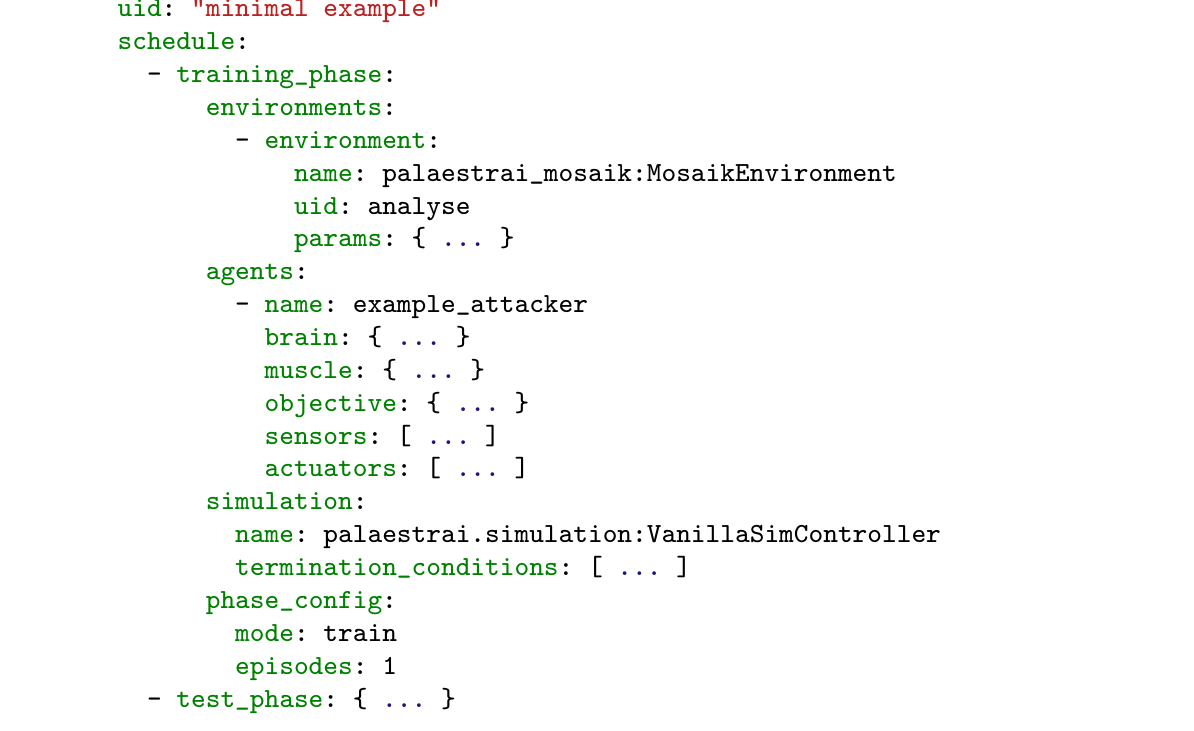}
    \caption{Simplified ANALYSE configuration file.}
    \label{listing:yaml_config_1}
\end{figure}
\par 

Such scenarios can be created in YAML files. Figure~\ref{listing:yaml_config_1} shows configuration snippets for our example. 
The core element is the schedule, which defines the different phases of the scenario, \eg, training and testing of the learning agents.
For each phase, configuration parameters for environments and agents need to be provided.
For the environment, this includes the configuration for the co-simulation scenario with \midas/mosaik.
The behavior of an \palaestrai agent is defined by brain and muscle, which together form an agent, objective and the available sensors and actuators. 


\section{Impact}\label{sec:impact}


The impact of \smasher can be derived directly from its features and the co-developed frameworks.
As mentioned before, most research in the field focuses on single domains and sub-systems. 
The coupling of power systems, markets, \ac{ICT}, and potentially other domains enables researchers to look at new research questions that focus on interconnected multi-domain systems and their interrelations. For example, \smasher has the potential to investigate research questions like "How much can profit on the energy market be increased when we apply false data injection at point xy in the system?" or "How much of the \ac{ICT} needs to be compromised by an attacker to create a blackout?". 
Normally, such research questions would require extensive implementation and modeling. With \smasher, the co-simulation only needs to be supplemented by the required simulators and models with little implementation effort, if they do not exist already.  
\par 
Second, we can place learning agents at almost arbitrary places in the system. By defining their reward function as maximizing damage, we have a general tool to investigate potential attack strategies in \ac{CPES}. Similarly, it is possible to find market exploits by defining market profit as reward \cite{wol21}. 
However, the opposite is imaginable too. By defining the reward function as maximizing, \eg, system stability or welfare, we can determine possible defensive strategies like controller settings by learning them. Therefore, it is also possible to place attacker \textbf{and} defender agents into the system to investigate their interrelations, which is the \ac{ARL} approach \cite{fischer2018adversarial}. However, the  main focus of \smasher is the attack and vulnerability analysis.
\par 
Besides enabling new research questions, \smasher will also prove helpful for existing research directions. Again, modularity and configurability are the key functionalities: 
For example, Wolgast \etal \cite{wol21} investigate attack scenarios in a coupled energy and reactive power market system. With \smasher, it would be possible to reproduce the research by removing the \ac{ICT} simulator and defining the \ac{RL} actuators, sensors, and reward according to the objective in the paper. Chen \etal \cite{che19b} apply false data injection attacks to attack automatic voltage control in the power network. Their work could be reproduced by removing the market simulation and again defining the \ac{RL} problem accordingly. 
In conclusion, \smasher builds a foundation to reproduce existing research in the field, investigate different variants of these scenarios, or increase their complexity.
\par 
\smasher is still newly published. However, we believe it can provide an important contribution to investigating unwanted attack scenarios in \ac{CPES} and to deriving potential countermeasures. 
\par 


\section{Conclusions}\label{sec:conclusion}
\smasher is a co-simulation-based tool-suite to use \ac{DRL} to find and analyze attack strategies in \aclp{CPES}. Its current version combines a power system, \ac{ICT}, and an energy market implementation. We designed \smasher to be modular, configurable, and self-documenting to allow for the investigation of diverse research questions. This way, \smasher is designed to reproduce a broad range of existing research and to find new attack strategies in interconnected multi-domain systems. 
\par 
Currently, \smasher focuses on a single learning attacker agent, but the long-term idea is to allow for multi-agent systems, add defender agents, analyze their interplay, and improve system design based on the findings. 




\section{Conflict of Interest}


We wish to confirm that there are no known conflicts of interest associated 
with this publication and there has been no significant financial support for 
this work that could have influenced its outcome.


\section*{Acknowledgements}
This  work  was  funded  by  the  German Federal Ministry of Education and Research through the research project PYRATE (01IS19021A).

\section*{Author contributions}

\textbf{Thomas Wolgast:} Conceptualization, Investigation, Methodology, Software, Validation, Writing - original draft, Writing - review \& editing

\textbf{Nils Wenninghoff:} Conceptualization, Investigation, Methodology, Software, Validation, Writing - review \& editing

\textbf{Stephan Balduin:} Conceptualization, Investigation, Methodology, Software, Validation, Writing - review \& editing

\textbf{Eric Veith:} Conceptualization, Investigation, Methodology, Software, Validation, Writing - review \& editing

\textbf{Bastian Fraune:} Conceptualization, Investigation, Methodology, Software, Validation, Writing - original draft, Writing - review \& editing

\textbf{Torben Woltjen:} Conceptualization, Investigation, Methodology, Software, Validation, Writing - review \& editing

\textbf{Astrid Nieße:} Conceptualization, Investigation, Methodology, Writing - review \& editing

\bibliographystyle{unsrt}  
\bibliography{references}

\end{document}